\documentclass[Proceedings]{ascelike}

\usepackage[left=2.5cm,top=2.5cm,right=2.5cm,bottom=2.5cm]{geometry}
\usepackage[utf8]{inputenc}
\usepackage[T1]{fontenc}
\usepackage{amsmath}
\usepackage{graphicx}
\usepackage{float}
\usepackage{parskip}
\usepackage[font=small,labelfont=bf]{caption}
\usepackage{epstopdf}
\usepackage{longtable}
\usepackage{lineno}

\begin{document}


\title{SENSITIVITY ANALYSIS AND STATISTICAL CONVERGENCE OF A SALTATING PARTICLE MODEL}

\author{Sergio Maldonado
\thanks{
PhD Candidate, Institute for Energy Systems, School of Engineering, The University of Edinburgh, The King's Buildings, Edinburgh EH9 3JL, U.K. E-mail: S.Mal-Vil@ed.ac.uk Tel: +44 (0)131 651 7293. Fax: +44 (0)131 650 6554} and Alistair G.L. Borthwick
\thanks{Professor of Applied Hydrodynamics, Institute for Energy Systems, School of Engineering, The University of Edinburgh, The King's Buildings, Edinburgh EH9 3JL, U.K. E-mail: alistair.borthwick@ed.ac.uk Tel: +44 (0)131 650 5588 Fax: +44 (0)131 650 6554}
}

\maketitle

\begin{abstract}
Saltation models provide considerable insight into near-bed sediment transport. This paper outlines a simple, efficient numerical model of stochastic saltation, which is validated against previously published experimental data on saltation in a channel of nearly horizontal bed. Convergence tests are systematically applied to ensure the model is free from statistical errors emanating from the number of particle hops considered. Two criteria for statistical convergence are derived; according to the first criterion, at least $10^3$ hops appear to be necessary for convergent results, whereas $10^4$ saltations seem to be the minimum required in order to achieve statistical convergence in accordance with the second criterion. 
Two empirical formulae for lift force are considered: one dependent on the slip (relative) velocity of the particle multiplied by the vertical gradient of the horizontal flow velocity component; the other dependent on the difference between the squares of the slip velocity components at the top and bottom of the particle. The former is found to give more stable results. A parameter study indicates that the saltation length has a minimum value with increasing particle diameter (at non-dimensional $D_* \sim 12$) for a given transport stage. Variations in the friction coefficient and collision line level have negligible effect on the saltation statistics within the ranges considered. Regression equations are obtained for each of the saltation characteristics. Finally, the model is used to evaluate the bed load transport rate, which is in satisfactory agreement with common formulae based on flume data, especially when compared against other saltation-derived expressions.\\

\end{abstract}


\section{Introduction}

Bed load transport consists of a combination of three different types of particle motion, namely: rolling, sliding and saltation. Whereas rolling and sliding occur near the threshold of incipient motion (for a recent study on initiation of motion see e.g. \citeNP{Diplas}), saltation only occurs once the particle motion has far exceeded the threshold \cite{vRijn1984}. Saltation is then related to a higher flow shear velocity, and hence tends to be responsible for most of the total bed load transport. 
For this reason, saltating particle models are often used to gain insight into bed load sediment transport. Such models, however, vary considerably from each other depending on which forces are taken to act on the grain and the approach adopted to the collision-rebound mechanism of the particle with the bed (or splash function) when continuous saltation is simulated (for single-hop models see e.g. \citeNP{vRijn1984}; \citeNP{Lee1994}). The most sophisticated equations of motion include effects due to turbulence, rotation of the particle (Magnus force) and changing boundary layer (Basset history term), at the cost of considerable mathematical complexity (see e.g. \citeNP{NG1998b}). With regard to the splash function, stochastic methods are commonly used due to the inherent randomness of the collision-rebound phenomenon (deterministic methodologies have also been developed, but these usually require highly idealised assumptions concerning the composition of the bed; see e.g. \citeNP{Yan2010}). For example, \citeN{Sekine1992} stated 
that the position of the particles' centroid forming the bed surface relative to the mean bed level followed a Gaussian distribution. However, \citeN{NG1994} assumed the bed was composed of uniformly packed spheres. Ni{\~n}o and García then defined a collision surface whose angle, with respect to the horizontal, was generated as a random number from a uniform distribution dependent on the incidence angle. In both the foregoing models, and after defining the friction and restitution coefficients, geometrical considerations were adopted to obtain the take-off conditions (i.e. magnitude and direction of particle velocity). For this class of stochastic models, it is usual to report the average value of the characteristics measured after a certain number of hops have been simulated. However, the number of hops is often set in a rather arbitrary fashion (e.g. \citeNP{NG1998b} simulate 400 hops, whereas \citeNP{Lee2000} choose only 55 saltations and \citeNP{Sekine1992} work with the range between 100 and 500 
realisations); and thus, there is no guarantee that statistically convergent results will be achieved.\\
\\
The model presented herein belongs to the stochastic-type of continuous-saltation models. However, it differs from others (e.g. \citeNP{Sekine1992}; \citeNP{NG1994}) in that no explicit assumptions are made regarding the structure of the bed. Instead, laboratory data are used to predict directly the take-off angle (by means of randomly generated numbers), hence leaving the friction coefficient as the sole calibration parameter to be determined in order to continue the saltation process. The main advantage of this alternative approach is the considerable simplification of the mathematical model and computer code required. The model is then used to develop two criteria aimed at estimating the minimum number of hops to be simulated in order to ensure statistically convergent results when similar saltating particle models are employed. The sensitivity of the model to variations in the friction coefficient (often considered constant), the type of lift force formula selected and the level at which the particle 
strikes the bed (herein referred to as the ``collision line'') is also assessed.\\ 
\\
The aim of the present paper is to propose a mathematically simple, computationally efficient stochastic model for fast, accurate simulation of saltating particles within the bed load transport zone. The model, herein referred to as Simple Saltating Particle (SSP) model, is then utilised to carry out a detailed study of statistical convergence and a sensitivity analysis. Section \ref{sec_model_descr} describes the mathematical model based on the equations of motion of a saltating particle and the collision-rebound mechanism or splash function. Section \ref{sec_validation} focuses on model validation against data available in the literature. The choice of lift force formula is also examined. Section \ref{sec_stat_convergence} discusses the two criteria derived for statistical convergence in order to ascertain the minimum number of particle hops required to be simulated. Section \ref{sec_sens_analysis} presents the results of parameter studies examining the effect of random variations in the values of bed friction coefficient and collision line level on the saltation 
statistics. In Section \ref{sec_regression_eqs}, regression equations are used in order to derive formulae for the different saltation characteristics. The model is used to compute the bed load transport rate in Section \ref{sec_bedload}, and is then compared against other formulae found in the literature. Section \ref{sec_conclusions} summarises the main conclusions and recommendations.\\
\\

\section{Model description} \label{sec_model_descr}
The model solves the equations of motion for a saltating spherical particle in combination with a stochastic approach in order to simulate the collision-rebound mechanism of a saltating particle with the bed, as explained below. (See Figure 1 for reference)

\subsection{Equations of motion}
The equations of motion of a saltating particle include contributions from inertia, lift, drag, and submerged weight. The Basset history term (related to the time-changing boundary layer on the particle) and the Magnus force (associated with the rotation of the particle) are not considered herein in order to retain model simplicity. Stochastic methods are used to account for the idealisations considered (e.g. Basset history term, Magnus force, random fluctuations in the time-averaged fluid velocity encountered in turbulent flows, irregularity in shape and layout of particles on the bed surface, etc.).\\

\begin{figure}[H] 
\centering
\fbox{\includegraphics[height=4cm]{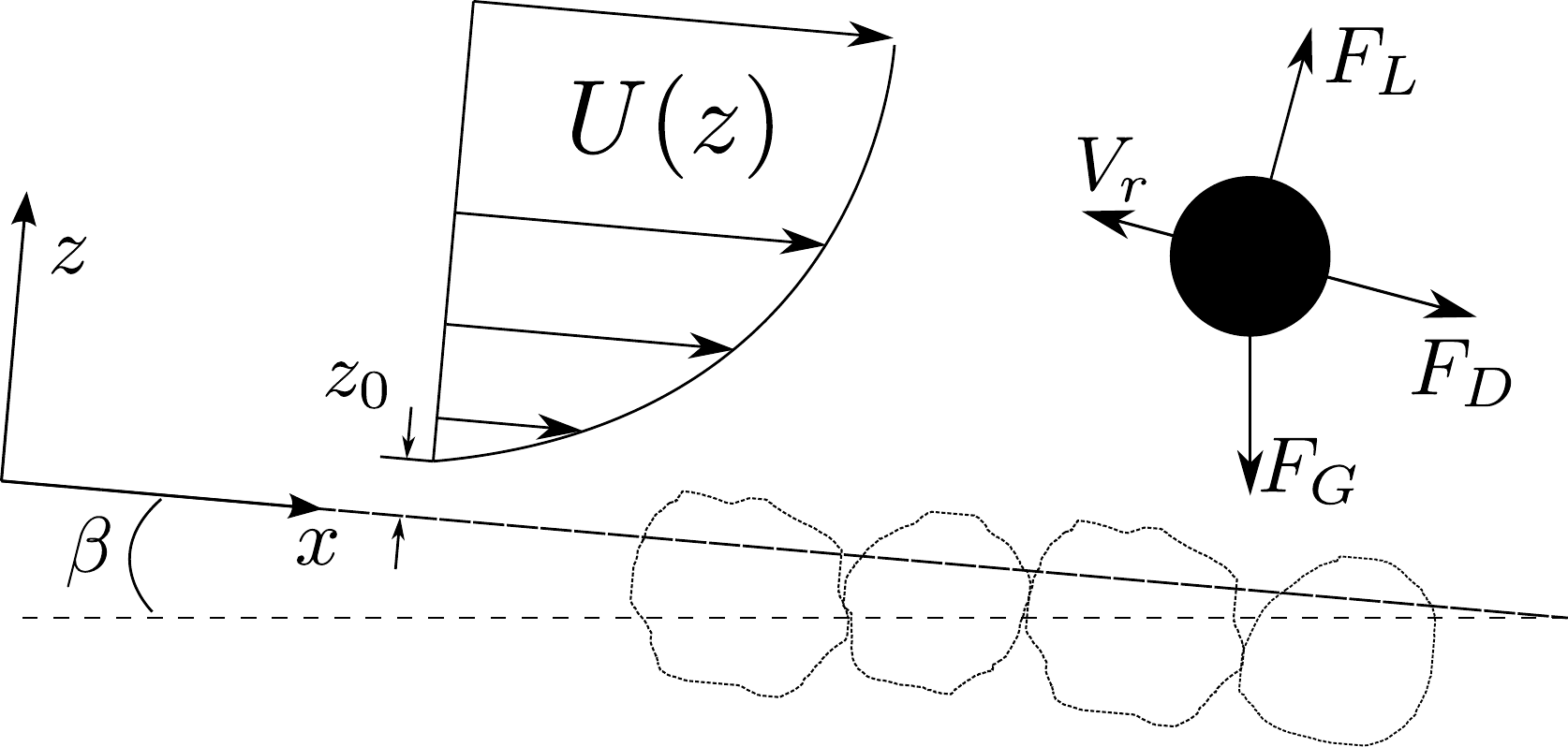}}
\caption{Definition sketch for the model}
\end{figure}

The governing equations are (see e.g. \citeNP{vRijn1984}; \citeNP{Lee1994}):

\begin{subequations} \label{eqs_motion}
 \begin{align}
 m \ddot{x} &= F_L \left (  \frac{\dot{z}}{V_r} \right ) + F_D \left ( \frac{U-\dot{x}}{V_r} \right ) + F_G \sin \beta \\
 m \ddot{z} &= F_L \left ( \frac{U-\dot{x}}{V_r} \right ) - F_D \left (  \frac{\dot{z}}{V_r} \right ) - F_G \cos \beta
\end{align}
\end{subequations}

where $x$ and $z$ are the streamwise and bed-normal centroid displacements of the particle; $m=(\rho_s + \alpha_m \rho) \pi D^3 / 6$ is the total mass of the particle with $\rho$ and $\rho_s$ being the densities of fluid (water in this case) and particle, respectively; $D$ is the particle diameter and $\alpha_m$ is the added mass coefficient, theoretically equal to 0.5;  $\beta$ represents the bed slope, having a (negative)positive value for (up)down-sloping beds, following the convention in river engineering; $F_L$, $F_D$ and $F_G$ are the lift, drag and gravitational force components, respectively; $U$ is the horizontal velocity component of the fluid; and $V_r$ is the slip or relative velocity of the particle evaluated at its centroid as follows:

\begin{equation} \label{rel_vel}
 V_{r}^{2} = (\dot{x}-U)^2 + \dot{z}^2
\end{equation}

The drag force, acting in the direction opposite to that of the relative velocity, is caused by a combination of form drag and skin friction \cite{Lee1994} and is expressed, for spherical particles, as:

\begin{equation}
 F_D = \frac{1}{8} C_D \;  \pi D^2 \; \rho \; V_{r}^{2}  
\end{equation}

where $C_D$ is the drag coefficient. However, the behaviour of $C_D$ for unsteady flow has not been widely analysed \cite{NG1998b} and so it is usual to estimate its value from that of a single, steady, free-falling particle. The empirical formula given by \citeN{Swamee} is employed herein, namely:

\begin{equation}
 C_D = 0.5 \left \{ 16 \left [ \left ( \frac{24}{R} \right )^{1.6} + \left ( \frac{130}{R} \right )^{0.72} \right ]^{2.5} + \left [ \left ( \frac{40,000}{R} \right )^{2} + 1 \right ]^{-0.25} \right \}^{0.25}
\end{equation}

where $R = \left | V_r \right | D / \nu$ is the particle Reynolds number and $\nu$ represents the kinematic viscosity of the fluid.\\
\\
Less is known about the lift force than the drag. Until few decades ago, the very existence of the lift force was questioned: \citeN{Bagnold1973} argued that the only upward impulses exerted on the particle were those due to collision with the bed. However, it is now widely accepted that there is enough evidence to state that hydrodynamic lift forces play a fundamental role in saltation \cite{NG1998b}. Here, two formulae are used to calculate the lift force, and the results compared.  The first was proposed by \citeN{Staffman} and used by \citeN{vRijn1984} in a single-hop saltating particle model, namely:

\begin{equation} \label{fl_vrijn}
 F_{L_1} = \alpha_L \rho \nu^{0.5} D^{2} \, V_r \left ( \frac{\partial U}{\partial z} \right )^{0.5}
\end{equation}

where $\alpha_L$ is the lift coefficient. Equation \eqref{fl_vrijn} was originally derived for small Reynolds numbers but later applied to the turbulent regime with a linearly varying $\alpha_L$ until it reached a maximum value \cite{vRijn1984}. Nonetheless, the lift coefficient $\alpha_L$ is herein used as a constant-value calibration parameter within the model (following the usual convention).\\  
\\
The second formula for $F_L$ comes from \citeN{Anderson} who considered wind-driven sediment transport, and is expressed for spherical particles as:

\begin{equation} \label{fl_lee}
 F_{L_2} = \frac{1}{8} C_L \rho \pi D^{2} ( V_{rT}^{2} - V_{rB}^{2} )
\end{equation}

where $V_{rT}$ and $V_{rB}$ are the relative velocities calculated using \eqref{rel_vel}, evaluating $U$ at the top and bottom of the particle, respectively. $C_L$ in the above equation represents the lift coefficient, which is used as a calibration parameter.\\
\\
The submerged weight is expressed as:

\begin{equation}
 F_G = \frac{1}{6} \pi (\rho_s - \rho) g D^{3}
\end{equation}

where $g$ is the acceleration due to gravity. The time-averaged vertical structure of the flow velocity is assumed to follow a logarithmic profile, given by:

\begin{equation}
 U(z) = \frac{U_*}{\kappa} \ln  (z/z_0)
\end{equation}

where $\kappa = 0.4$ is the von Kármán constant; $z_0 = 0.11 (\nu / U_*) + 0.033 k_s$ is the bed level at which velocity is zero; $U_*$ is the shear velocity and $k_s$ is the equivalent roughness height of Nikuradse, taken to be proportional to the size of the bed material. It is important to mention that the use of the above velocity logarithmic profile assumes a low concentration of particles in the bed load area (when relating the mechanics of a single saltating particle to the study of bed load transport), which agrees with experimental findings (e.g. \citeNP{FernandezLuque}); for higher concentrations the effect of particles on the fluid vertical structure would not be negligible.\\
\\
Equation \eqref{eqs_motion} is transformed into a system of first order ordinary differential equations and then integrated in time using a fourth-order Runge-Kutta method.\\

\subsection{Collision-rebound mechanism (splash function)} \label{sec_collision-rebound}

Consider the collision-rebound event depicted in Figure 2. From the equations of motion the velocity vector at the moment of collision, $\vec{V}_{in} = (\dot{x},\dot{z})_{in}$, is determined. For simplicity, it is assumed that $\vec{V}_{in}=\vec{V}_{n}$; in other words, equal to that immediately before the event. As the vertical position at collision is known (the level $z_{in}$ defines the collision line), the time needed for the particle to reach that line can be obtained as: $t_{in} = (z_n - z_{in})/ \dot{z}_n$, and the horizontal position of collision as: $x_{in} = \dot{x}_{in} t_{in} + x_n$. Notice that $t_{in} < \Delta t$, where $\Delta t$ represents the numerical time step. The point $(n+\Delta t)_{virtual}$ in Figure 2 represents the point predicted by the solution of the governing equations for which the condition $z < z_{in}$ is first verified, i.e. when the collision line is virtually ``crossed''. The take-off angle is then generated as a random number based on 
measurements reported in the literature (see e.g. \citeNP{NG1998}; \citeNP{Lee2000}). Herein this angle is assumed to follow a normal distribution, i.e. $\theta_{out} \sim N(\mu,\sigma^2)$, where $\mu$ and $\sigma^2$ are the angle mean value and variance, respectively. The striking horizontal velocity, $\dot{x}_{in}$, is thought to be reduced after collision by a factor $f$, allowing therefore the calculation of the take-off streamwise and vertical velocities as $\dot{x}_{out} = f \, \dot{x}_{in}$ and $\dot{z}_{out} =  \dot{x}_{out} \tan \theta_{out}$, respectively. Notice that the restitution coefficient, defined as $e = -\dot{z}_{out} / \dot{z}_{in}$, is herein a result of the splash function rather than a tuning parameter. An alternative approach would be to set $e$ as the tuning variable, leading to $f \; (=\dot{x}_{out} / \dot{x}_{in})$ being the resultant variable instead. However, it has been found from laboratory data that, unlike the restitution coefficient, the friction coefficient exhibits fairly 
constant behaviour \cite{NG1998} and thus it is chosen to be the main calibration parameter within the collision model.\\

\begin{figure}[H] 
\centering
\fbox{\includegraphics[height=4cm]{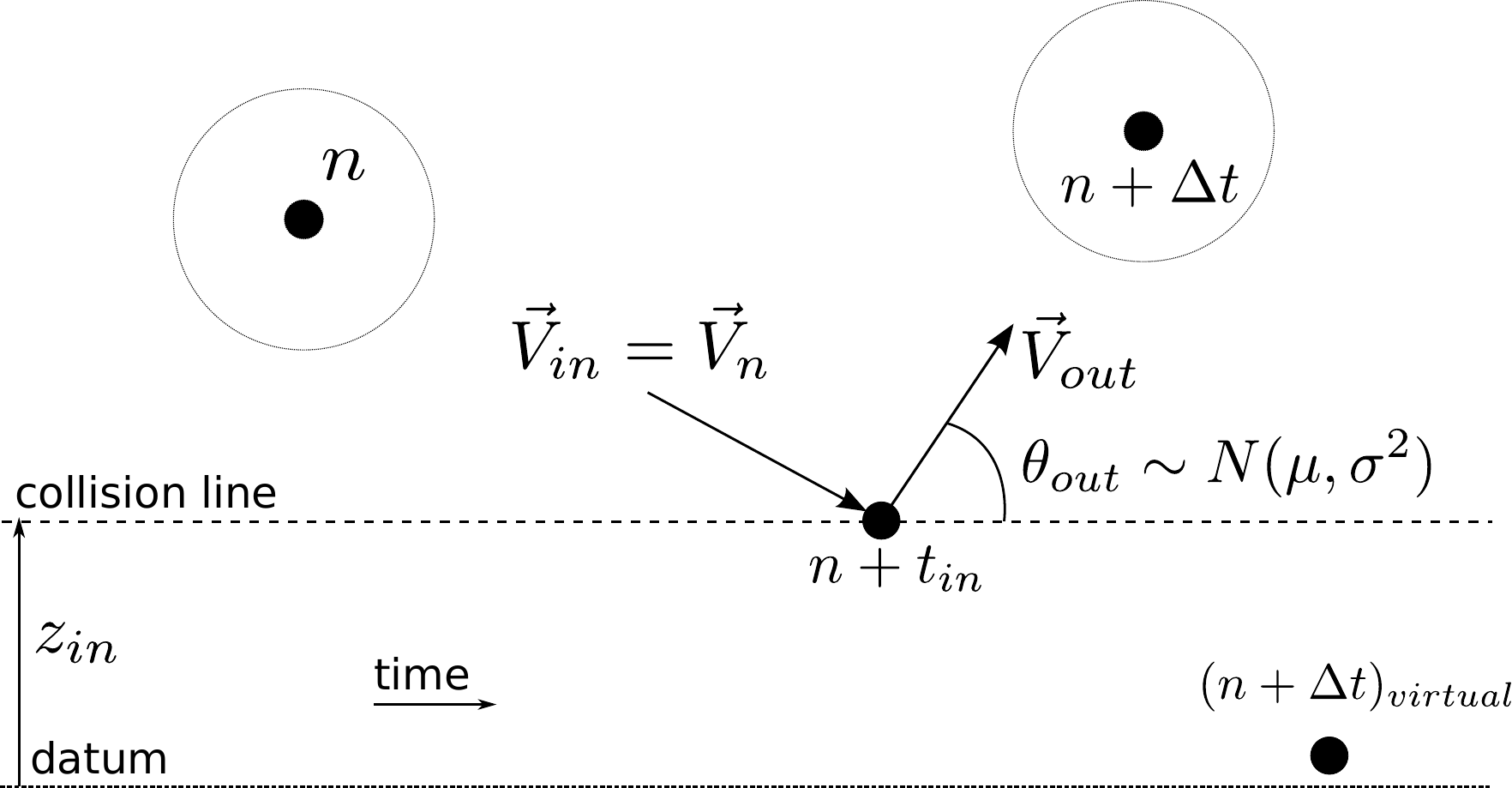}}
\caption{Sketch of the splash function} \label{fig_collision}
\end{figure}

It is worth mentioning that attention has to be paid to the definitions of $e$ and $f$ adopted, especially when compared to other studies. For instance, \citeN{NG1998} define the coefficients as reductions in the tangential and vertical velocity components with respect to the collision surface, which does not necessarily coincide with the stream-wise plane (as in the present model). It should therefore not be expected that the values of $e$ and $f$ should be the same. For model validation, the friction coefficient, $f$, and the position of the collision line, $z_{in}$, may be considered constant. This simplification is revised in Section \ref{sec_sens_analysis}.\\
\\
Once the take-off velocity, $\vec{V}_{out} = (\dot{x},\dot{z})_{out}$, is obtained, the new position of the particle is found by linear interpolation as follows: 

\begin{subequations}
 \begin{align}
  x_{n+\Delta t} &= x_{in} + \dot{x}_{out} (\Delta t - t_{in}) \\
  z_{n+\Delta t} &= z_{in} + \dot{z}_{out} (\Delta t - t_{in})
 \end{align}
\end{subequations}

The post-rebound velocity is assumed to be equal to the take-off one; i.e. $\vec{V}_{n+\Delta t} = \vec{V}_{out}$. From this point on, the numerical integration of the governing equations continues as normal until the particle again encounters with the collision line.\\
\\
It is important to highlight that the present model does not replicate the exact behaviour of a saltating particle, such as the eventual rest to which a particle comes when trapped by the local bed topology. Hence the term ``bed surface'' is deliberately avoided and the term ``collision line'' is used instead. Following the idea behind Monte Carlo simulation, continuous saltation of the particle is modelled until statistical convergence of the sampled characteristics is achieved. The main goal of the model is to evaluate these average characteristics.
The results presented in this paper should then be interpreted as probabilistic tendencies arising from a statistical analysis of saltation (based on a combination of laboratory data and relatively simple governing equations), rather than as findings resulting from strictly physics-based numerical simulations.\\
\\

\section{Validation} \label{sec_validation}

Model predictions (SSP $F_{L_1}$ and SSP $F_{L_2}$) are compared against the experimental results reported by \citeN{Francis1973} (F), \citeN{Lee1994} (LH), \citeN{NG1994} (NG A 1 and NG A2, where the datasets differ from each other with respect to the diameters measured) and \citeN{NG1998} (NG). The characteristics measured include the saltation height, $\delta_s$, length, $\lambda_s$, and stream-wise velocity of the particle, $u_s$, as depicted in Figure 3, which illustrates a typical trajectory followed by the centroid of a saltating particle. The conditions replicated in the simulations are those of the experiments by \citeN{NG1998}. A particle diameter of 0.5 mm is used and values of the shear velocity, $U_*$, in the range from 0.0207 to 0.0321 m/s are modelled. During the comparison, the transport stage, defined as $\tau_* / \tau_{*c}$, is used as the independent variable (the abscissa in Figures 4 to 6). $\tau_*$ denotes the dimensionless bed shear stress evaluated as $\tau_* = U_{*}^{2} / [g(s-1)D]$, where $s = \rho_s / \rho$ is the sediment relative density; $\tau_{*c}$ represents the 
dimensionless critical shear stress for sediment motion, obtained from the Shields curve. Note that the values of $\tau_* / \tau_{*c}$ herein computed do not match those of \citeN{NG1998}. The reason is that Ni{\~n}o and García report variations in the measured particle size and therefore $\tau_*$ and $\tau_{*c}$, which are functions of $D$, are affected.\\

\begin{figure}[H] 
\centering
\includegraphics[height=4.5cm]{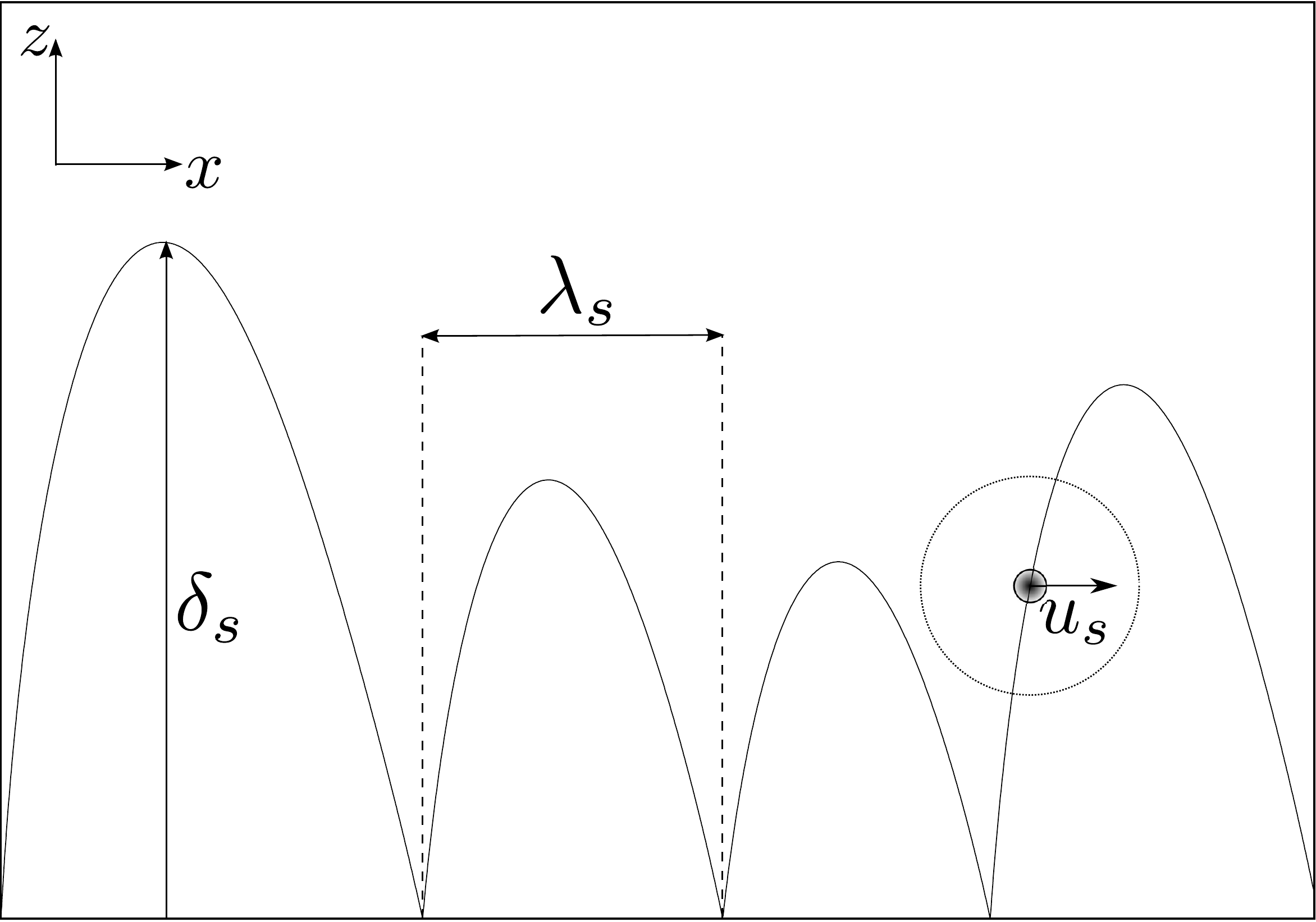}
\caption{Typical saltation trajectory and characteristics measured} \label{fig_typical}
\end{figure}

As input data the following values are used: $D = 0.5 \ \mathrm{mm}$, $\rho = 1,000 \ \mathrm{kg / m^3}$, $\rho_s = 2,650 \ \mathrm{kg / m^3}$, $\nu = 1.1 \times 10^{-6} \ \mathrm{m^2 / s}$, $k_s = 1 \, D$ and $\beta = 0.05 \ ^\circ$ (down-slope). $\theta_{out}$ is generated as a random number following a normal distribution with parameters $\mu = 25^{\circ}$ and $\sigma \sim 14^{\circ}$ ($\sigma^2 = 200$). The values of $\mu$ and $\sigma$ are estimated from the work of \citeN{NG1998} and \citeN{Lee2000}. Following \citeN{vRijn1984}, the collision line is set to be $0.6 \, D$ above the datum, i.e. $z_{in} = 0.6 \, D$. Two variables are used as calibration parameters: the lift coefficient, $\alpha_L | C_L$ (depending on the formula for lift force employed), and the friction coefficient, $f$. Three hundred continuous hops are simulated for every scenario. This number is of a similar order to those commonly reported in experiments and it is sufficient to make the influence of arbitrary initial conditions 
negligible, herein set to $x=0$, $z=0.6 \, D$ and $\dot{x}=\dot{z}=2.5 \, U_*$ \cite{vRijn1984}. Section \ref{sec_stat_convergence} presents further discussion on the adequate number of hops to be simulated. First, the friction coefficient is considered to be constant; later the influence of randomness is also tested (see Section \ref{sec_sens_analysis}). From the calibration process, the optimum values of the tuning parameters are found to be $\alpha_L=10 \ | \ C_L=0.75$ and $f=0.85$. This value of $f$ is in perfect agreement with the findings of \citeN{FernandezLuque}.\\

\begin{figure}[H]
\centering
\includegraphics[height=7.5cm]{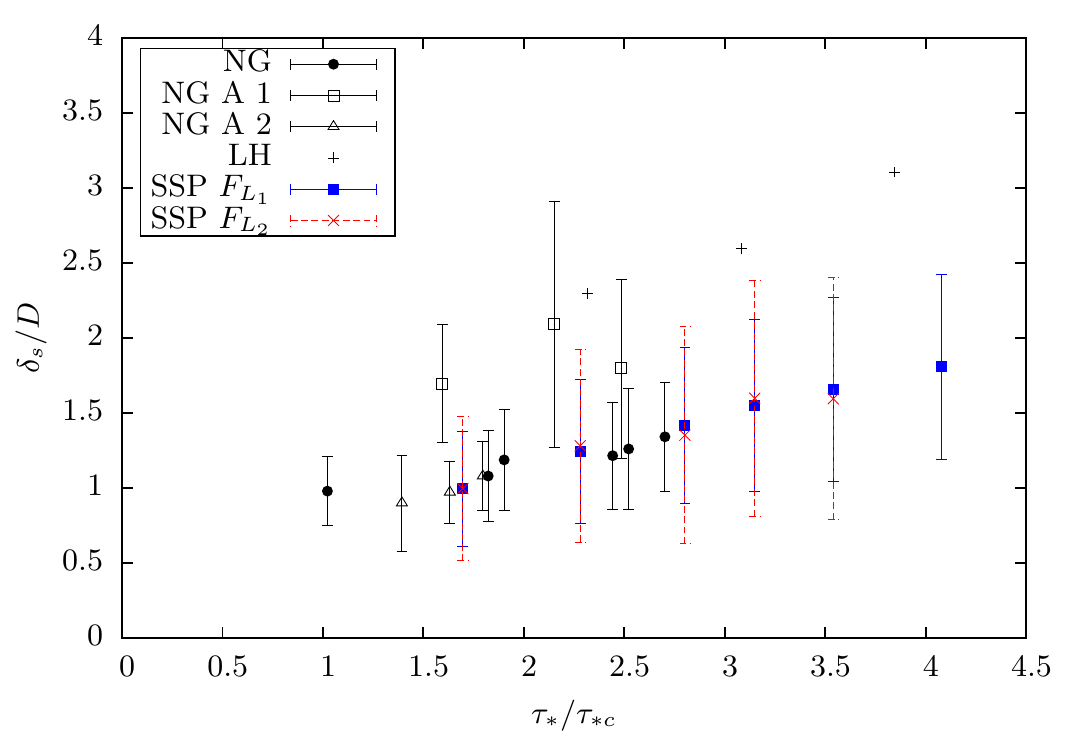}
\caption{Dimensionless saltation height versus transport stage [Symbols represent mean values and vertical lines represent the total length of two standard deviations; acronyms in legend defined in corresponding paragraph]} \label{fig_validation_sh}
\end{figure}

\begin{figure}[H]
\centering
\includegraphics[height=7.5cm]{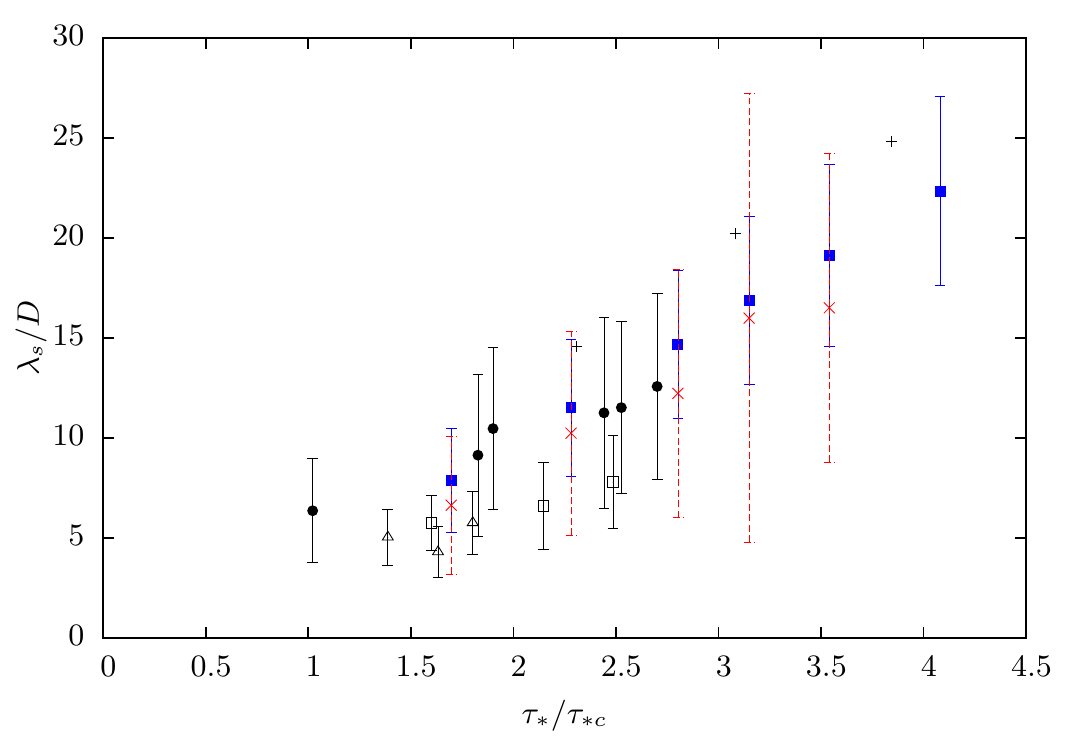}
\caption{Dimensionless saltation length versus transport stage [Symbols, vertical lines and legend as in Figure 4]} \label{fig_validation_sl}
\end{figure}

\begin{figure}[H]
\centering
\includegraphics[height=7.5cm]{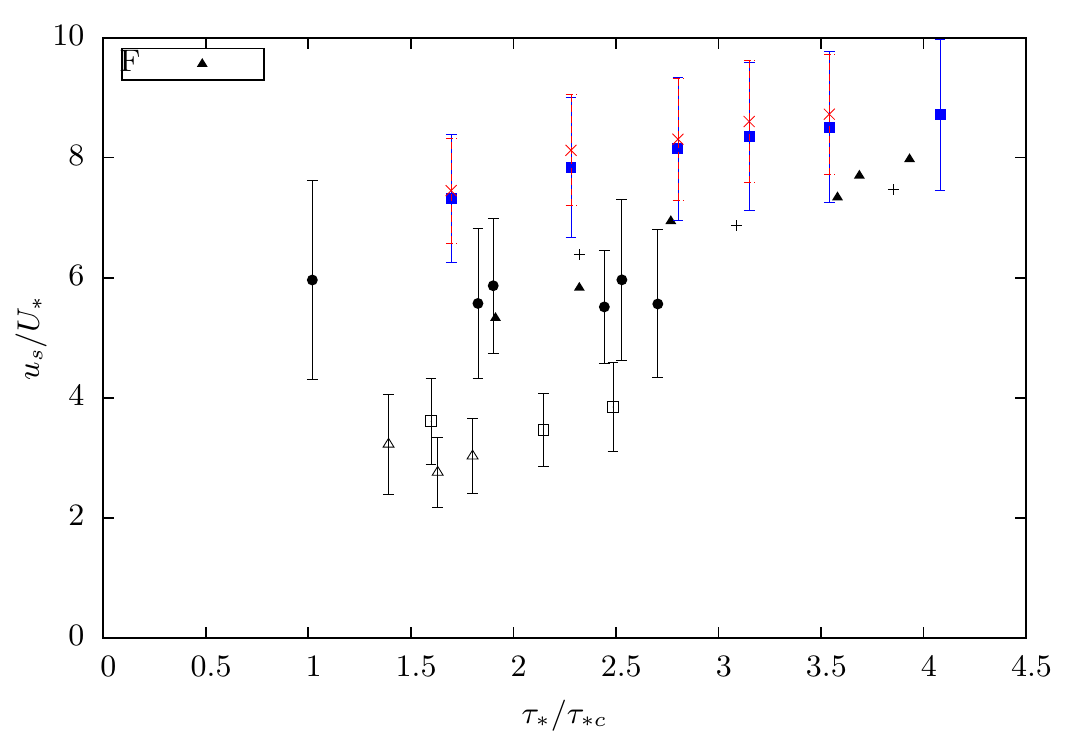}
\caption{Dimensionless saltation stream-wise velocity versus transport stage [In legend, F denotes \protect\citeN{Francis1973}; the remaining symbols are as in Figures 4 and 5]} \label{fig_validation_us}
\end{figure}

Figures 4 and 5 show the non-dimensional saltation height and length as functions of the transport stage. Figure 6 illustrates the corresponding plot of non-dimensional saltation stream-wise velocity (made dimensionless with the shear velocity). The definition of saltation height given in the open literature is often ambiguous; here the definition stated by \citeN{NG1998} is adopted, i.e. the maximum distance reached during a hop between the centroid of the particle and the top of the bed grains. The top of the bed grains is assumed to be $0.25 \, D$ above the datum \cite{vRijn1984}. This assumption underpins the values depicted in Figure 4. 
The model shows generally good agreement (both in mean values and standard deviations) with the experimental data considered, particularly those of \citeN{NG1998} whose experiments are simulated in this work. Overestimations of $u_s$ can be perceived, which may be related to the formula used to compute the drag coefficient; however the model seems to predict well the asymptotic behaviour of this variable for increasing flow velocity, in accordance with previous work (see e.g. \citeNP{FernandezLuque}). Comparing the formulae for the lift force (points sharing the exact same values of $\tau_* / \tau_{*c}$ in the figures), two main conclusions arise: i) except for the case of $u_s$, $F_{L_2}$ shows a larger scatter (standard deviation) than $F_{L_1}$ for the same simulation conditions; and ii) when using $F_{L_2}$ the model eventually becomes unstable for increasing flow velocities, which has made it impossible to obtain results for $\tau_* / \tau_{*c} > 4$. Due to this limitation, hereafter $F_{L_1}$ will be used within the model.\\ 
\\

\section{Study of statistical convergence} \label{sec_stat_convergence}

The basic idea behind Monte Carlo methods is the repeated sampling of random (formally pseudo-random) numbers in order to simulate complex systems. These methods vary considerably, but conventionally, $m$ random numbers are sampled from a given probability distribution (input variables) and a deterministic computation is carried out in order to obtain the solution to the problem (output of the system). This process is repeated $n$ times until statistically convergent results are achieved. In the present work, this is translated as follows: $n$ numerical experiments are executed, each consisting of $m$ hops performed by the particle. Thus, for each experiment $m$ values of the take-off angle (the only variable so far defined as random) are generated. The deterministic algorithm applied to the input random variable is the solution to the equations of motion defining the path of the particle between the rebound and eventual collision with the bed. Notice that because of the nature of this 
problem, an equivalent approach to the $n$ experiments of $m$ hops each is to carry out a single numerical experiment simulating $m \times n$ hops (this deduction has been numerically verified). As stated in the previous section, the main outputs of the system are the saltation characteristics, i.e. particle saltation height, length and stream-wise velocity. In order to assess the statistical convergence of the results two criteria are adopted, following the methodology described below.\\
\\
A reference scenario is obtained by simulating a fairly large number ($10^6$) of hops, and different $n$-hops runs are compared against this case. The first convergence criterion is defined as the percentage error in the mean values in relation to the reference case, as follows:

\begin{equation} \label{eq_error}
\textup{error} = \left | \frac{X-X_{\textup{ref}}}{X_{\textup{ref}}} \right | \times 100
\end{equation}

where $X$ can be the average value of any measured characteristic (i.e. $\lambda_s$, $\delta_s$ or $u_s$) for a given number of hops simulated, and the subscript ``ref'' denotes the reference scenario (i.e. the $10^6$-hops simulation, in this case).\\
\\
Another way of evaluating statistical convergence is by assessing how close or far a given sample is from a well-defined probability distribution. Thus the second criterion compares (in a rather qualitative fashion) the probability density obtained from the diverse $n$-hops cases against the one obtained from the reference scenario.\\
\\
Five different numbers of hops have been simulated, varying from $10^2$ to $10^6$ increasing every intermediate order of magnitude. This is repeated for three values of the transport stage (TS), namely: $\tau_* / \tau_{*c} = $ 2.8, 4.1 and 10 (TS 1, TS 2 and TS 3, respectively); in order to analyse the influence of increasing flow velocities. The grain diameter considered in the validation of the model is adopted, $D = 0.5$ mm. Figure 7 summarises the results. The saltation height has been selected as the target variable for clarity and because it is representative of the convergence behaviour followed by the other variables. In order to avoid confusion, hereafter $\delta_s$ is defined as the maximum height reached by the centroid of the particle during a hop in relation to the collision line, as illustrated in Figure 3. 
Points with vertical lines represent the mean values and standard deviations of $\delta_s$ divided by the reference average value, $\delta_{s_{\textup{ref}}}$ , for ${\tau_* / \tau_{*c} = 4.1}$ (left $y$-axis). In order to provide further insight, Figure 7 also includes three curves showing the percentage error (right $y$-axis) corresponding to different transport stages. The lower part of Figure 7 shows the probability density of the averaged non-dimensional saltation height for different numbers of hops modelled for $\tau_* / \tau_{*c} = 4.1$. In all cases, the Freedman-Diaconis rule has been used in order to set the number of bins. Several features arise from analysis of Figure 7. From a practical point of view it can be noted that mean values are usually relatively close to the reference case, with errors no larger than 10\% ; also, that this error decreases to less than 1\% if $10^3$ or more hops are simulated. The reference case shows a well defined gamma-type distribution that starts to be clearly recognisable after $10^4$ hops. Hence, based on the first convergence criterion it can be argued that $10^3$ hops seem sufficient to assure convergence; however, in accordance with the second criterion, $10^4$ hops appear to be better. This is interesting as most studies on continuous saltation report results for simulations of the order of hundreds of hops or less (e.g. \citeNP{NG1998b}; \citeNP{Lee2000}; \citeNP{Sekine1992}). Furthermore, the magnitude of the standard deviation seems to be independent of the number of hops.\\ 

\begin{figure}[H] 
\centering
\fbox{\includegraphics[width=\linewidth]{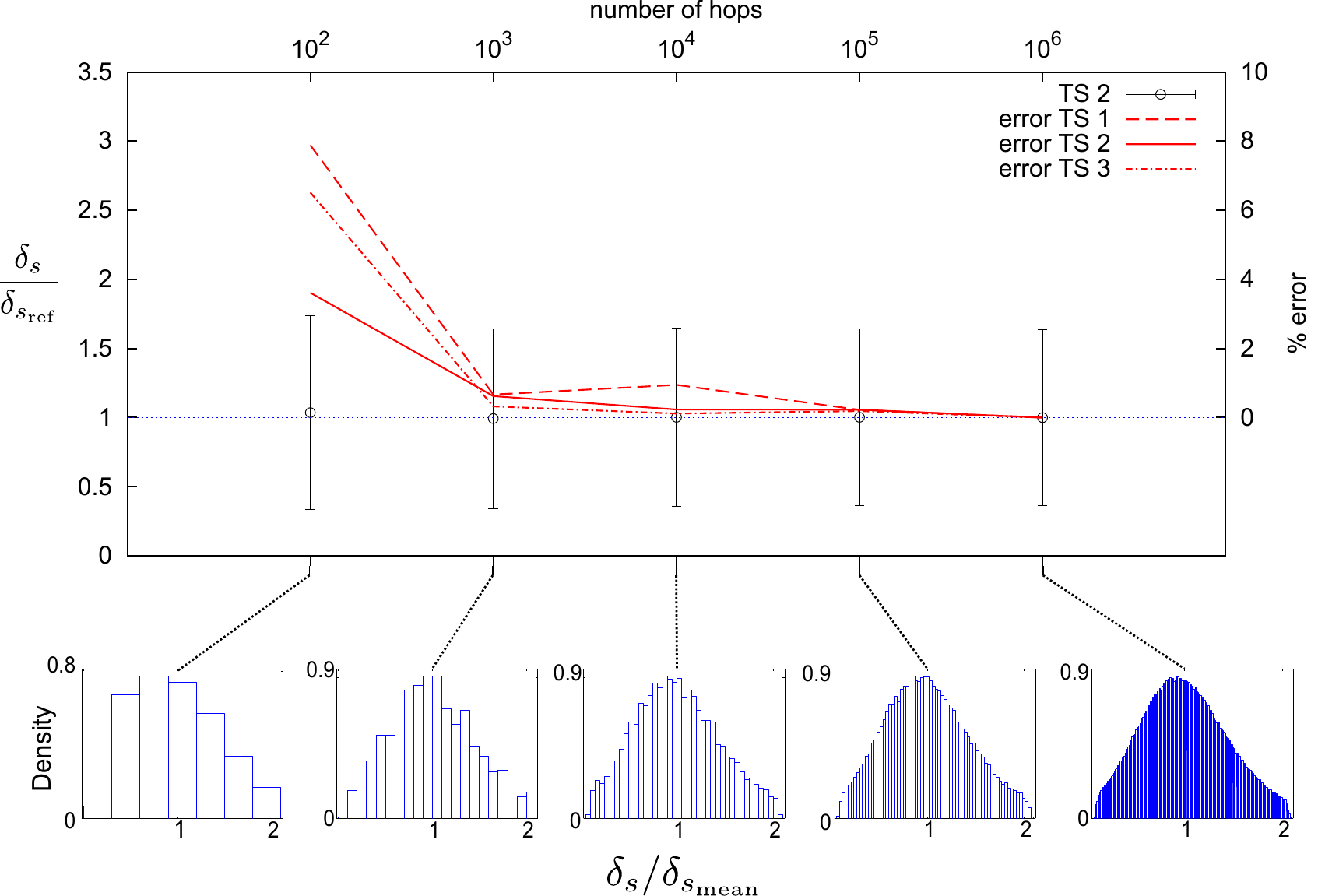}}
\caption{First and second convergence criteria, corresponding to upper and lower parts of the figure, respectively} \label{fig_converge}
\end{figure}

Figure 8 illustrates how the measured $\delta_s$ tends towards the mean value (i.e. the standard deviation decreases) for increasing flow velocity conditions. Hence, it can be concluded that as the flow velocity increases it progressively influences the saltation process by reducing the scatter in saltation characteristics due to the (highly random) collision-rebound phenomenon. In other words, for increasing flow velocities the fluid becomes the dominant agent in the saltation process, minimising the influence that the random particle collision and rebound with the bed has on the deviation from the mean value of the saltation characteristics (i.e. it reduces such deviation). 
The relationship between the variance found in $\delta_s$ and the variance set to generate the input variable $\theta_{out}$ has also been studied. For values of the variance of $\theta_{out}$ in the range of 200-400 \cite{NG1998}, a linear response in the variance of the non-dimensional saltation height occurs in the range of 0.28-0.35 (i.e. standard deviation in the range of 0.53-0.59), which further confirms that the deviations of this variable are relatively independent of the other input parameters (i.e. number of hops and variance of $\theta_{out}$). Figures 7 and 8 focus on saltation height for the reasons explained above, and also because i) $u_s$ does not exhibit considerable scatter; and ii) $\lambda_s$ is not as relevant as $\delta_s$ when studying the bed load transport within conventional approaches.\\

\begin{figure}[H]
\centering
\includegraphics[height=6.0cm]{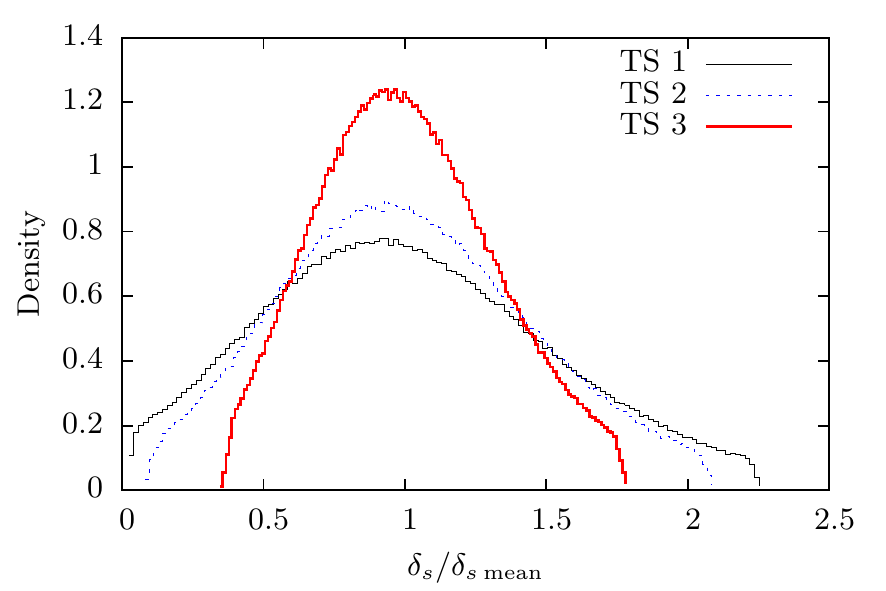}
\caption{Probability densities of $\delta_s$ divided by the mean values for different transport stages. Thin-solid, dashed and thick-solid lines correspond to TS 1, 2 and 3, respectively [labels as in Figure 7].} \label{fig_distributions_sh}
\end{figure}

Figure 9 depicts the probability distributions followed by the three main saltation characteristics after one million hops; for illustration, the case of $\tau_* / \tau_{*c} = 4.1$ (TS 2) is considered. Notice that, of the three variables, $\delta_s$ exhibits the largest deviation from the mean value, whereas values of $u_s$ tend to concentrate around the mean. A distribution-fitting test confirmed that the probability density of the saltation characteristics is closest to the gamma-type family (particularly to the Nakagami distribution). This is in accordance with the findings of \citeNP{Lee2000} for the case of $\delta_s$ and $\lambda_s$; Lee et al. concluded that these characteristics followed a Pearson Type III distribution; and that the measured(simulated) $u_s$ followed a normal(uniform) distribution. Lee \textit{et al.}'s results also show simulated values of $u_s$ being concentrated around the mean.\\

\begin{figure}[H] 
\centering
\includegraphics[height=6.0cm]{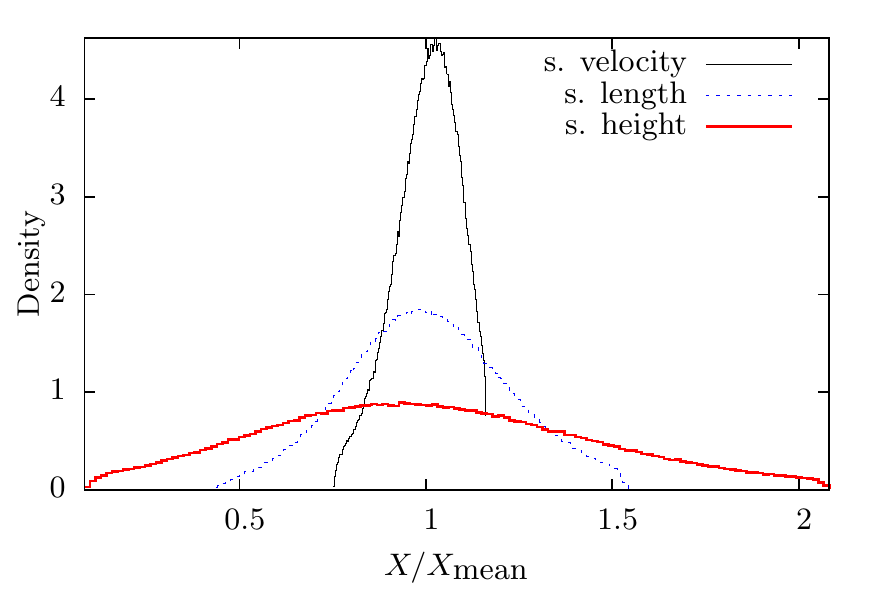}
\caption{Probability densities of the saltation characteristics divided by the mean values, for the case of TS 2. Thin-solid, dashed and thick-solid lines correspond to the saltation stream-wise velocity, length and height, respectively.} \label{fig_distributions}
\end{figure}

This study of statistical convergence also demonstrates an important feature of the present model: its mathematical simplifications permit a fast computer program to be used in order to simulate a large number of hops in an efficient manner. As an illustration, 10,000 hops can be simulated in about 1 minute (depending on the value of the shear velocity modelled) using a standard PC (i.e. Intel Core i3 3.10 GHz processor, 4 GB RAM). This feature is particularly important in the studies carried out in the following sections.\\
\\

\section{Sensitivity analysis} \label{sec_sens_analysis}

A constant value of $f$ (which is the usual approach) implies that the stream-wise reduction of momentum is constant (i.e. collision-rebound events always occur in the same way). However, even though data reported in literature show an arguably constant behaviour of this coefficient, unsurprisingly, they also exhibit clear scatter (see e.g. \citeNP{NG1998}). For this reason, the possibility of having a variable friction coefficient is now tested, by generating $f$ from a uniform probability distribution, such that $f \sim U(f_{min},f_{max})$. Hence, $f_{min}$ and $f_{max}$ denote the minimum and maximum values of $f$ permitted, respectively. In a similar fashion to $f$, the influence of randomness on the collision line level, $z_{in}$, is also evaluated. This is done with the aim of taking into account the diverse irregularities inherent to the grains within the near-bed area (e.g. size, position, shape, etc.), both while saltating and resting on the bed, and which directly affect the position of the collision 
line. 
These two factors have been selected for the sensitivity analysis given their direct impact on the collision-rebound mechanism (considered herein as the main goal of investigation within the saltation phenomenon). Sensitivity of the model to other factors so far neglected (such as Magnus force, turbulent fluctuations in the velocity profile, etc.) would provide a more profound insight into the understanding of a saltating particle; however, such analysis would probably merit another piece of work for future research.\\
\\
By giving $f$ and $z_{in}$ (which were previously treated as constants) values associated with a probability distribution, the influence of randomness is assessed from the effect it has on the mean values and standard deviations of the saltation characteristics, as well as on their statistical convergence. The Base Case is taken where both $f$ and $z_{in}$ are constant, as previously validated. Three different combinations are tested: friction coefficient being constant(variable) with a variable(fixed) collision line, and both variables being random.\\
\\
Case 1: constant $f$, random $z_{in}$. Whereas the friction coefficient has a calibrated constant value, i.e. $f=0.85$, the value of $z_{in}$ is generated from the uniform distribution, ranging from a lower limit of $0.6 \, D$ \cite{vRijn1984} up to a value of $0.75 \, D$. These limits are obtained from geometrical considerations, as depicted in Figure 10.\\
\\
Case 2: random $f$, constant $z_{in}$. For the reasons discussed above, the friction coefficient is generated as a uniformly distributed random number. Based on the available published literature, the value of this coefficient is allowed to vary within $\pm 0.1$ from the calibrated constant value, hence $f \sim U(0.75,0.95)$. The collision line is given as a constant value, fixed at $0.6 \, D$ above the datum.\\
\\
Case 3: random $f$, random $z_{in}$. In this case, the combined influence of randomness on the two variables is tested. The values of $f$ and $z_{in}$ are generated as described above.\\

\begin{figure}[H] 
\centering
\includegraphics[height=3.5cm]{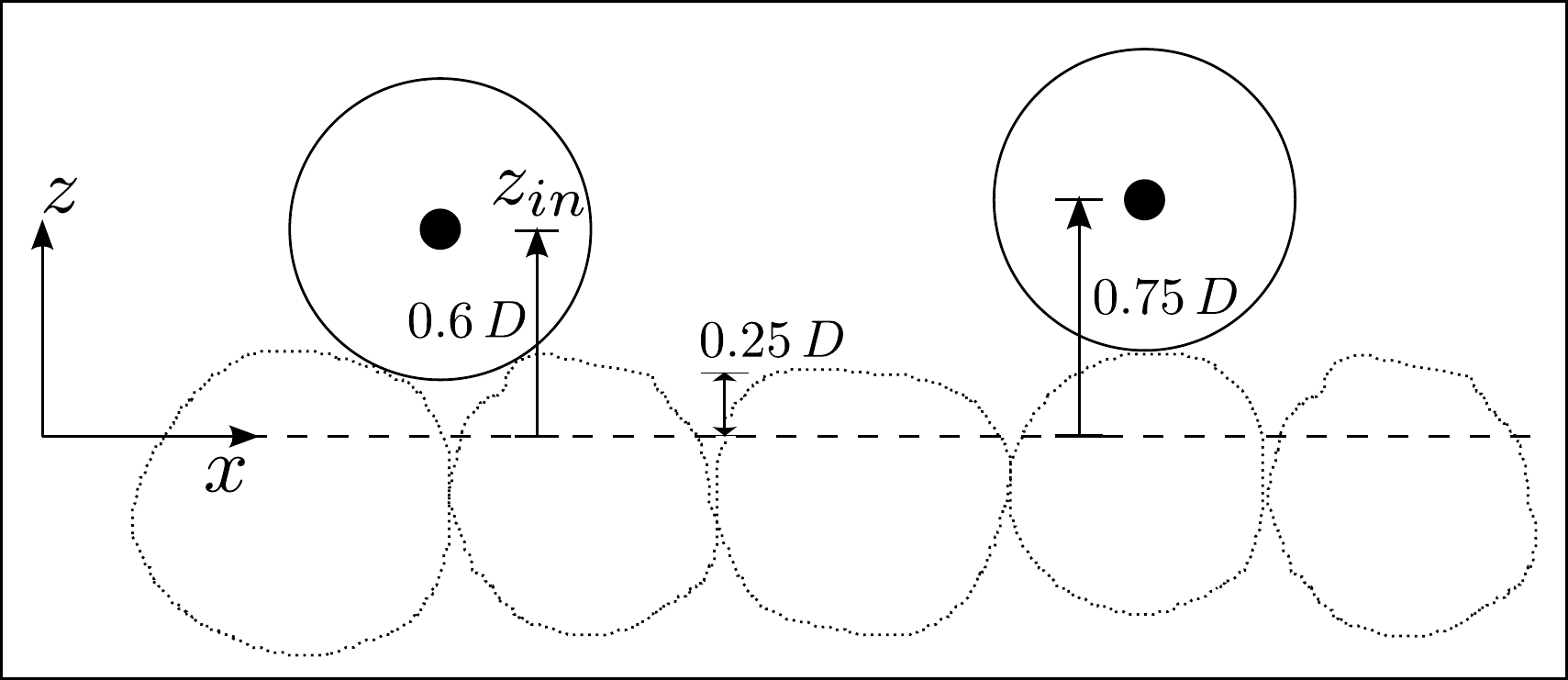}
\caption{Sketch illustrating lower and upper limits of a variable collision line}
\end{figure} \label{fig_random_zin}

The same methodology described in Section \ref{sec_stat_convergence} is applied herein to study statistical convergence. A reference scenario of $10^6$ hops is simulated for each case and results from different $n-$hops runs are compared against the reference results. Mean values and standard deviations of the three saltation characteristics, obtained after $10^6$ hops, are compared for each case against the Base Case (where both $f$ and $z_{in}$ are constant). The transport stage modelled is $\tau_* / \tau_{*c} = 4.1$. Figure 11 illustrates the percentage changes in mean and standard deviation of saltation height, length and velocity for each case, with respect to the Base Case. It can be observed that for all cases, the variation in the mean values falls within $\pm 2.5$\% with respect to the Base Case. Regarding the standard deviations, variations of $\delta_s$ and $u_s$ range from approximately -1 to 4\%; however, a larger increase is present in the standard deviation 
for $\lambda_s$, up to values of about 12 and 15\% for Cases 2 and 3, respectively, demonstrating the direct impact that a variation of the friction coefficient (reduction of the stream-wise velocity) has on the stream-wise distance reached by the particle during a hop (i.e. saltation length).\\

\begin{figure}[H] 
\centering
\includegraphics[height=6.5cm]{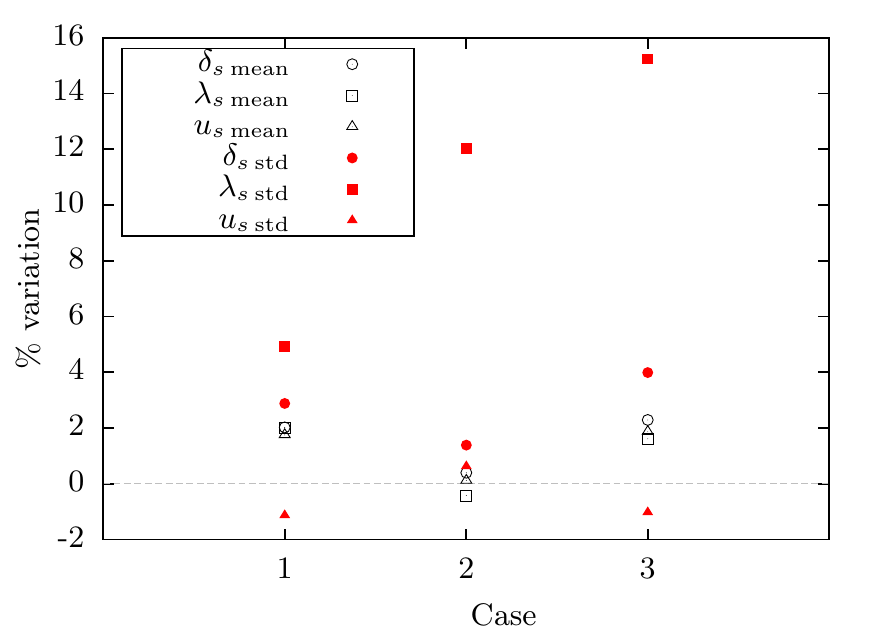}
\caption{Percentage variation of the saltation characteristics for each case in relation to the Base Case (i.e. $f$ and $z_{in}$ constant). Unfilled symbols represent mean values; filled points denote standard deviations} \label{fig_changes_mean_std}
\end{figure}

The first criterion for statistical convergence is depicted in Figure 12, where the percentage error of $\delta_s$, calculated from eq. \eqref{eq_error}, for each case is presented as a function of the number of hops. The Base Case (curve TS 2 in Figure 7) is also included. Observe that errors for $10^6$ hops are not plotted because these are equal to zero by definition. The second convergence criterion is shown in Figure 13, which illustrates the probability density of the saltation height obtained for each case and the number of hops simulated. The saltation height and transport stage are representative of the behaviour of the other saltation characteristics and values of $\tau_* / \tau_{*c}$ regarding statistical convergence. Figures 12 and 13 confirm the conclusions found in Section \ref{sec_stat_convergence} regarding statistical convergence, namely: at least $10^3$ hops have to be simulated in 
order to assure a mean value of the characteristics with an error smaller than 1\% (first criterion); and $10^4$ hops or more should be modelled so that the probability distribution followed by the resultant saltation features resembles sufficiently the final gamma-type distribution obtained after $10^6$ hops (second criterion). Figure 13 also shows an interesting feature of the final distribution resulting from Cases 2 and 3 (middle and bottom most-right panels): a small peak in the density as $\delta_s / \delta_{s_{\textup{mean}}}\rightarrow 0$. Given that Cases 2 and 3 both have uniformly distributed friction coefficients, this peak seems to occur when a small value of $f$ coincides (in a collision event) with a small value of $\theta_{out}$, therefore resulting in a small take-off velocity (and hence limited height reached by the particle). This peak is related to the joint probability of occurrence of simultaneously small values of both $f$ and $\theta_{out}$. Furthermore, it 
should be noted that, as in the Base Case, the deviation from the mean value (i.e. the magnitude of the standard deviation) is independent of the number of hops for the three cases studied. Overall, it can be concluded that the influence of randomness on $f$ and $z_{in}$ is negligible when focusing on the mean values and convergence behaviour presented by the saltation characteristics.\\

\begin{figure}[H] 
\centering
\includegraphics[height=6cm]{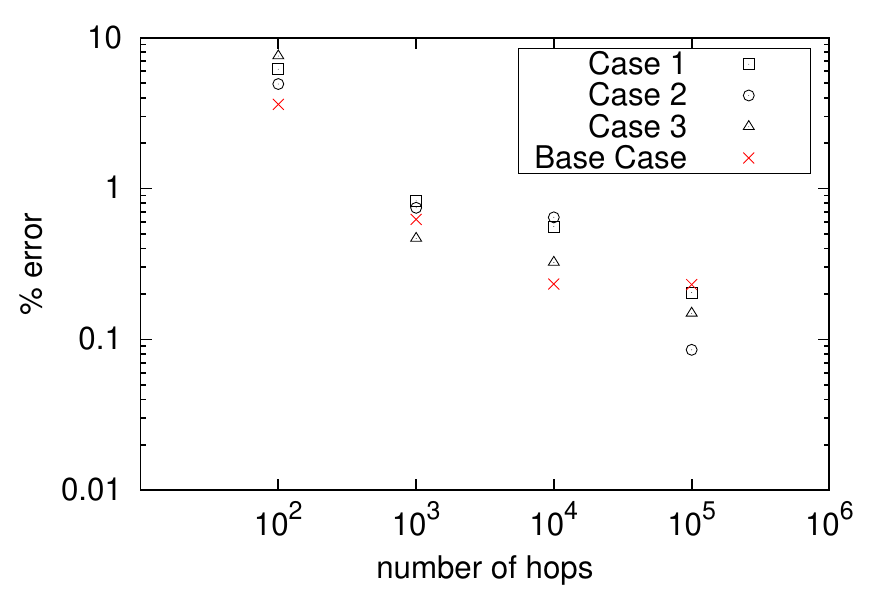}
\caption{First convergence criterion for the three cases and the Base Case, for $\delta_s$} \label{fig_conv_parameters}
\end{figure}

\begin{figure}[H] 
\centering
\fbox{\includegraphics[width=\linewidth]{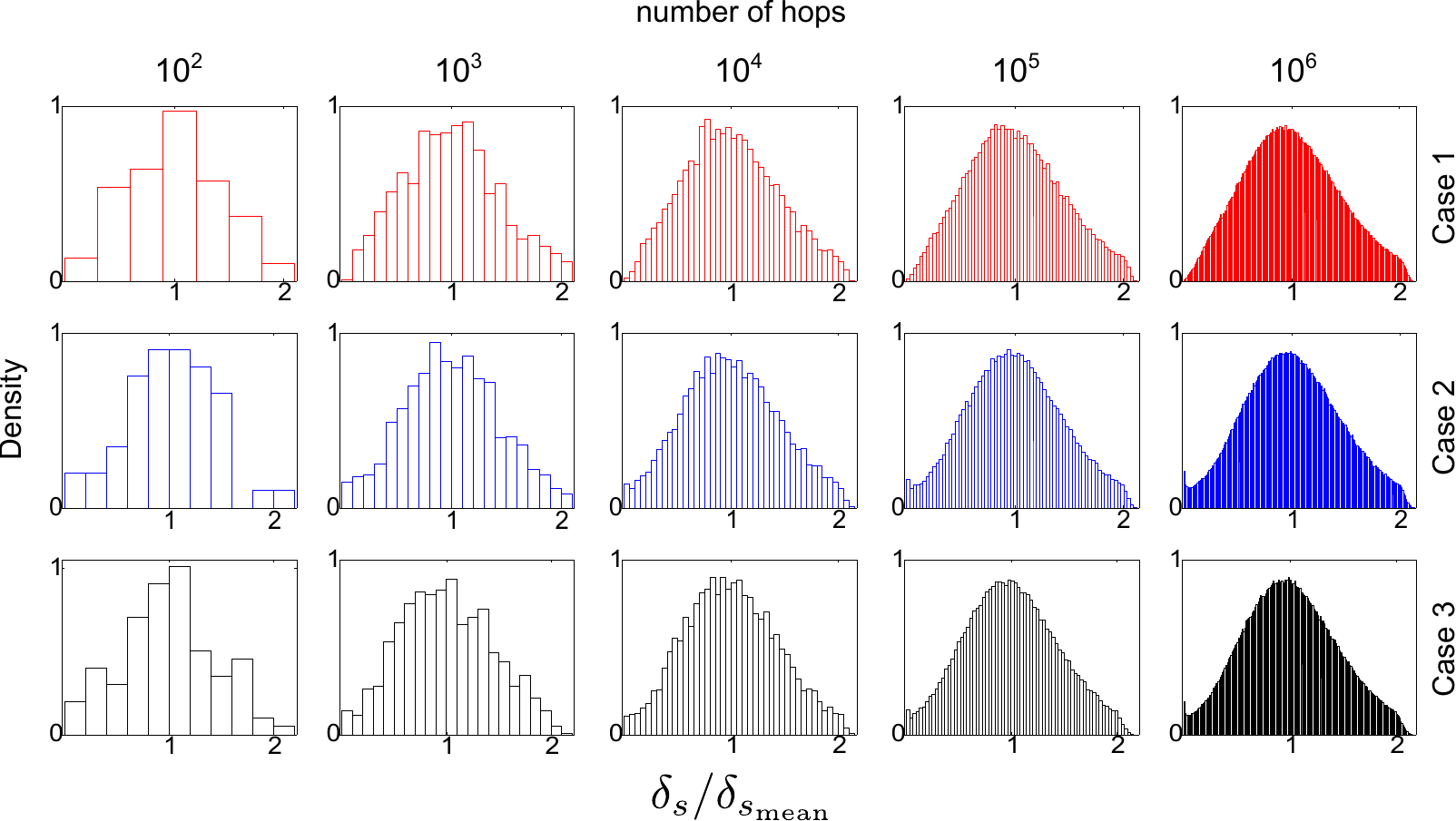}}
\caption{Probability densities of $\delta_s$ divided by the mean value (second convergence criterion) for Cases 1, 2 and 3 (top to bottom) and different numbers of hops simulated} \label{fig_2crit_parameters}
\end{figure}

\section{Regression equations} \label{sec_regression_eqs}

90 combinations are simulated, including 6 different particle sizes ($D=$0.1, 0.25, 0.5, 1.0, 2.0 and 4.0 mm) and values of $\tau_* / \tau_{*c}$ in the range of 1 to 16. In accordance with the convergence criteria previously developed, $10^4$ hops are simulated for each combination of $D$ and $\tau_* / \tau_{*c}$ in order to compute the mean value of the saltation characteristics. Regression equations are then obtained for the three saltation characteristics. Based on the work of other authors (see e.g. \citeNP{Lee2000}; \citeNP{vRijn1984}), equations of the form $X_* = a \, D_{*}^{b} \, T_{*}^{c}$ are adopted; where $X_*$, $D_*$($\equiv  D [(s - 1)g / \nu^2]^{1/3}$) and $T_*$($\equiv \tau_* / \tau_{*c}$) denote the non-dimensional saltation characteristic, diameter and transport stage, respectively; $a$, $b$ and $c$ are coefficients. For the saltation velocity, the form $u_s/U_* = a + b \ln D_* + c T_{*}^{-0.5}$ is also tested.\\
\\
The saltation length exhibits behaviour not predicted by other studies, to the authors' knowledge. Figure 14 presents the variation in $\lambda_s / D$ with $D_*$ and $T_*$. It can be seen that, for a given $T_*$, the profile of $\lambda_s / D$ evolves with $D_*$ to drop rapidly to a minimum value at $D_* = 11.87$ (corresponding to $D = 0.5$ mm in this case); from that point on, it grows in an asymptotic fashion. Note that this value of the diameter corresponds to the minimum critical shear stress in the Shields’ curve. When instead plotted against shear stress (the figure not included here for brevity), the saltation length progressively increases with both $D_*$ and  $\tau_*$. This underlines the importance of selecting an adequate independent variable ($\tau_*$ vs $T_*$) in studies of particle saltation. For this reason, a step function is herein adopted in the regression equation for $\lambda_s$. The other 
saltation characteristics do not present similar behaviour. However, it should be noted that in practice most approaches to the bed load transport using saltating particle models disregard the saltation length; typically, the bed load transport is modelled as the product $\delta_s u_s c_b$, where $c_b$ represents the sediment concentration within the bed load layer. A comment on this is made in the next section.\\

\begin{figure}[H] 
\centering
\includegraphics[height=4.5cm]{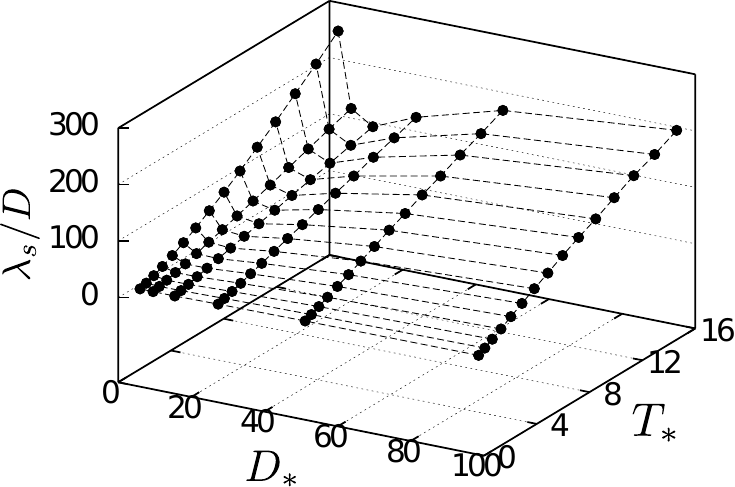}
\caption{Simulated non-dimensional saltation length versus dimensionless diameter and transport stage} \label{fig_3d_sl}
\end{figure}

The regression equations obtained, valid for $T_* \ge 1$, are:

\begin{equation} \label{eq_regSH}
\delta_s / D = 0.154 \, D_{*}^{0.495} \, T_{*}^{0.73}
\end{equation}

\begin{equation} \label{eq_regSL}
\lambda_s / D = \left\{\begin{matrix}
11.25 \, D_{*}^{-0.62} \, T_{*}^{1.3} & \textup{if} \; D_*<12\\ \\ 
2.327 \, D_{*}^{0.35} \, T_{*}^{1.03} & \textup{if} \; D_* \geq 12
\end{matrix}\right.
\end{equation}

\begin{subequations} \label{eq_regSV}
\begin{align}
u_s / U_* &= 4.355 \, D_{*}^{0.14} \, T_{*}^{0.19} \label{eq_regSV1} \\
u_s / U_* &= 8.328 + 1.328 \ln D_* - 6.232 T_{*}^{-0.5} \label{eq_regSV2}
\end{align}
\end{subequations}

The corresponding values of the correlation coefficient $R'^2$ for equations \eqref{eq_regSH}, \eqref{eq_regSL}, \eqref{eq_regSV1} and \eqref{eq_regSV2} are 0.98, 0.99, 0.93 and 0.94, respectively. Care should be taken when comparing against other equations in the literature, given that different definitions of the transport stage, $T_*$, may be used. For example, another conventional definition of the transport stage is: $(U_{*}^{2} - U_{*c}^{2})/U_{*c}^{2}$ (where $U_{*c}$ is the critical shear velocity for initiation of motion), which is equal to $T_* - 1$, with $T_*$ as defined herein (i.e. $T_* \equiv \tau_* / \tau_{*c}$).\\
\\

\section{Bed load transport} \label{sec_bedload}

The study of saltation is very useful when attempting to understand the mechanics of bed load transport. However, when saltating particle models are used in order to compute the bed load sediment transport rate, $q_b$, several considerations have to be taken into account. For instance, at low flow regimes (near the threshold of motion), rolling and sliding may be prominent modes of transport and so a saltating-particle-derived formula for $q_b$ could underestimate the bed load transport rate under those conditions. On the other hand, at higher flow velocities, more particles are expected to entrain motion. However, if the number of grains in saltation is sufficiently large, results from an analysis like the present one may lose validity due to the influence that a large number of particles in saltation may have on the fluid velocity and the effect of inter-particle collisions. Nonetheless, the present model is utilised in order to calculate bed load transport and compare it against other formulae available in the literature. Figure 15 depicts a comparison between the non-dimensional bed load transport, $\Phi $($\equiv q_b / [g (s-1) D^3]^{1/2}$), calculated using the present model against the saltating-particle-derived formulae (SP) of \citeN{Lee2000} (L) and \citeN{vRijn1984} (vR), and the commonly used flume-data-derived expressions of \citeN{MeyerPeterMuller} (MPM), \citeN{Soulsby1997} (S) and \citeN{FernandezLuque} (FLB). A grain diameter of 2 mm with $\rho_s = 2,650 \ \mathrm{kg / m^3}$ has been used in all calculations. 
The relatively large discrepancies between the saltation-based expressions of \citeN{Lee2000} and \citeN{vRijn1984} and the flume-data-based formulae, may be due to the observations stated above in this paragraph, the definition of bed load layer, and the consequent estimation of its sediment concentration. The present model yields a more satisfactory agreement with the formulae by \citeN{MeyerPeterMuller}, \citeN{Soulsby1997} and \citeN{FernandezLuque} (in comparison with the other two saltation-based formulae), by computing the bed load transport as $q_b = h_b c_b u_s$; where the bed load concentration, evaluated as $c_b = 0.117 (T_* - 1)/D_*$ \cite{vRijn1984}, is taken as the sediment concentration of a bed load layer defined by the thickness $h_b$, related in turn to $\delta_s$ (calculated from eq. \ref{eq_regSH}) through idealised geometrical considerations as $h_b = \delta_s + D$ (see Figure 16). For the sake of simplicity, $u_s$ is evaluated using \eqref{eq_regSV1} instead of \eqref{eq_regSV2}. The curve calculated with the present model has been plotted up to a value of $\tau_*$ roughly corresponding to $c_b = 0.25$, which has been experimentally found to be the average maximum concentration in a bed load layer \cite{Spinewine2005}.\\

\begin{figure}[H] 
\centering
\includegraphics[height=7.5cm]{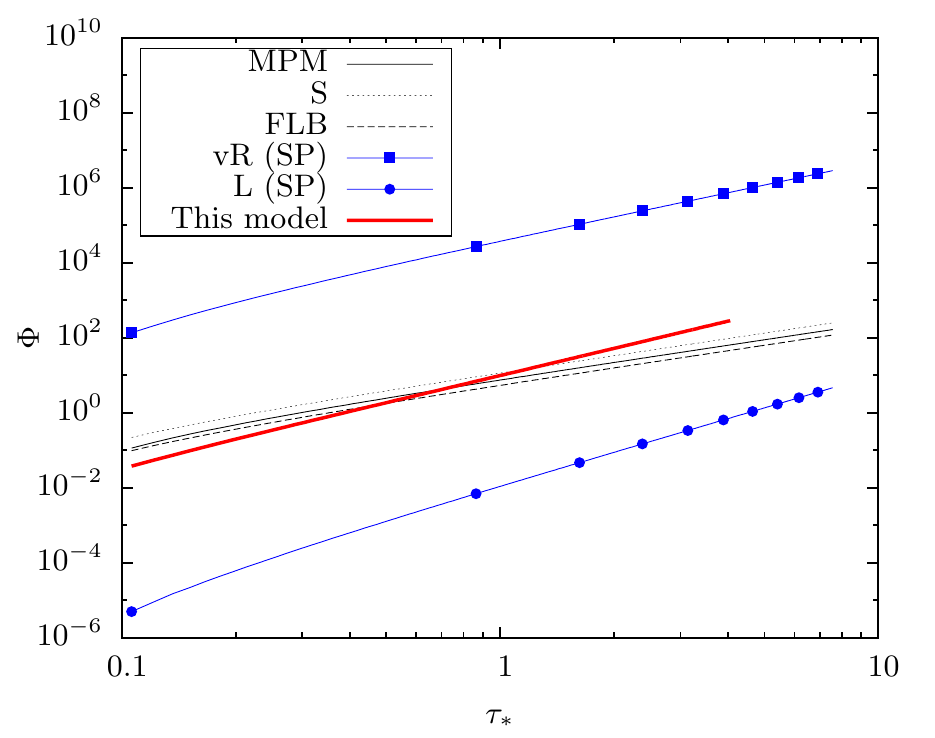}
\caption{Non-dimensional bed load transport versus non-dimensional bed shear stress [Acronyms in legend defined in corresponding paragraph].\\} \label{fig_bedload}
\end{figure}  

\begin{figure}[H] 
\centering
\fbox{\includegraphics[height=4.0cm]{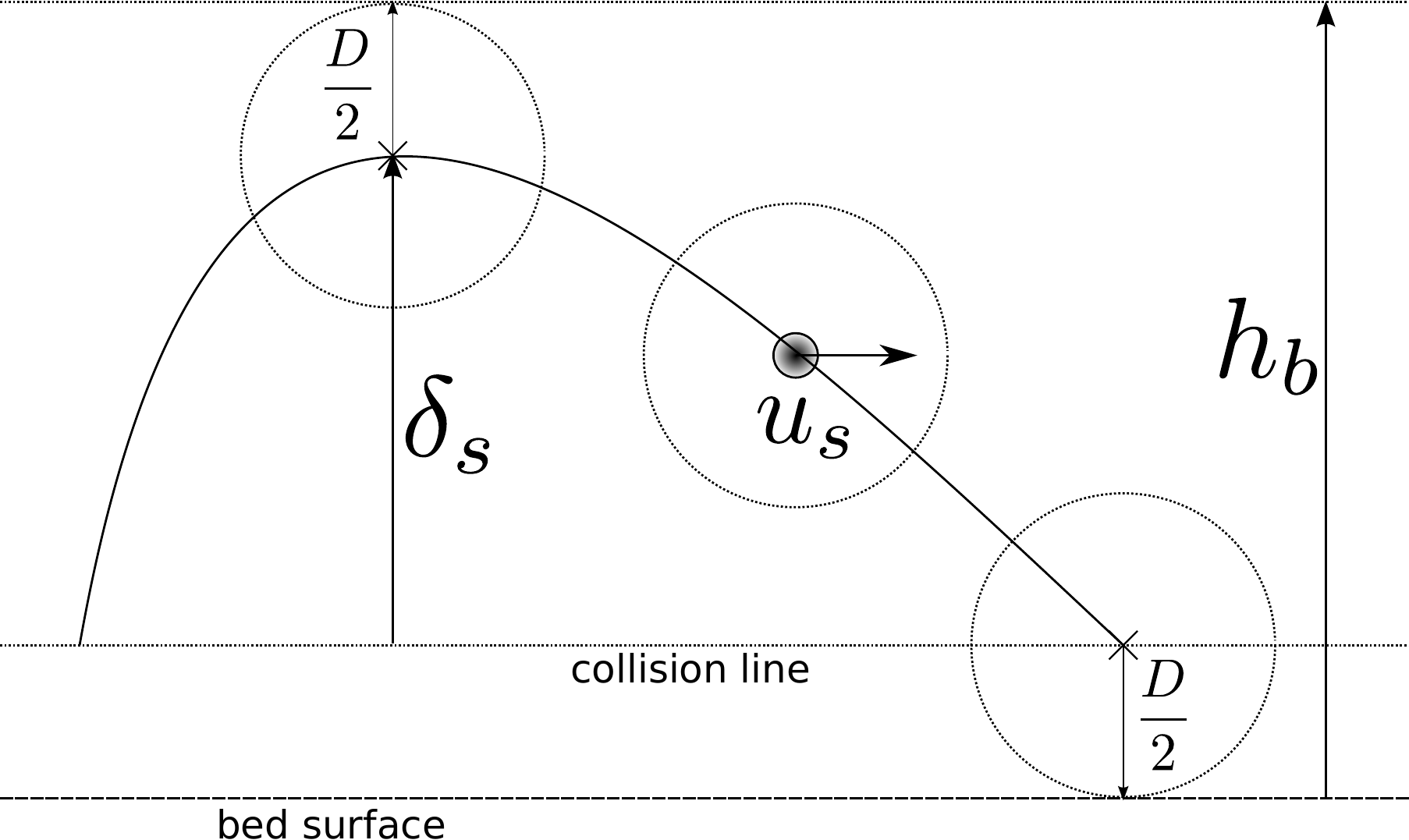}}
\caption{Sketch depicting relation between bed load layer thickness and particle saltation height} \label{fig_hb}
\end{figure}

\section{Conclusions} \label{sec_conclusions}

A fast, efficient numerical model for stochastic saltation has been developed utilising a simple splash function and governing equations, and validated satisfactorily against experimental data on saltation height, length, and velocity previously reported in the literature by  \citeN{Francis1973}, \citeN{Lee1994}, \citeN{NG1994} and \citeN{NG1998}. Two criteria for statistical convergence were identified: one related to the error in the mean values of the saltation characteristics between different $n$-hops runs and a large-$n$-hops (i.e. $10^6$ hops) reference scenario; the other concerned with the deviation in the sampled characteristics from a well-defined probability distribution achieved after a large number of hops (i.e. $10^6$) has been simulated. Model convergence tests show that at least $10^3$ particle hops are needed to satisfy the first criterion, and at least $10^4$ particle hops are required to satisfy the second criterion. This finding is relevant, given that some previous studies report 
results after only a few hundred particle hops or less have been simulated. 
The choice of empirical formula for the lift force component is obviously important – the present work has shown that a formula dependent on the slip (relative) velocity of the particle multiplied by the vertical gradient of the horizontal flow velocity component (i.e. $F_{L_1}$, see eq. \ref{fl_vrijn}) gives more stable results than a formula dependent on the difference between the squares of the slip velocity components at the top and bottom of the particle (i.e $F_{L_2}$, see eq. \ref{fl_lee}). A sensitivity analysis has shown that variations in the bed friction coefficient and the position of the collision line have almost no effect on the mean values and convergence behaviour presented by the saltation characteristics. 
The saltation height and velocity both increase monotonically with increasing particle diameter. The saltation length is also dependent on the particle diameter for a given value of $\tau_* / \tau_{*c}$, but with a minimum at a critical value of the non-dimensional particle diameter ($D_* \sim 12$,  corresponding to a particle diameter of about 0.5 mm for the test cases considered herein). This confirms the importance of the selection of an appropriate variable ($\tau_*$ vs $T_*$) when analysing the saltation characteristics. Regression analysis has been used to determine empirical formulae for the saltation height, length and stream-wise velocity of a particle over a nearly horizontal bed. The model has been used to compute the bed load transport rate, which is in good agreement with the commonly used formulae by \citeN{MeyerPeterMuller}, \citeN{Soulsby1997} and \citeN{FernandezLuque} (especially when compared against the saltation-derived expressions proposed by \citeNP{Lee2000} and \citeNP{vRijn1984}). Some remarks are made regarding the use of saltating particle models for calculation of bed load transport. In future, it would be useful to integrate a full Lagrangian model of saltation, rolling and sliding, to generate a more complete and insightful representation of bed load transport.\\
\\
\section{Acknowledgements}

The first author is supported by the Mexican National Council for Science and Technology (CONACYT) through Scholarship No. 310043.  The authors would also like to thank University College Cork, Ireland, where they were based previously and where much of the work reported herein was undertaken.\\
\\

\bibliography{SSPpaper_submit}

\section{Notation List}
\emph{The following symbols are used in this paper:}
\nopagebreak
\par
\begin{longtable} {r  @{\hspace{1em}=\hspace{1em}}  l}
$c_b$                    & bed load sediment concentration; \\
$C_D$                    & drag coefficient; \\
$C_L$                    & lift coefficient for $F_{L_2}$; \\
$D$                    & particle diameter; \\
$D_*$                    & non-dimensional particle diameter; \\
$e$                    & restitution coefficient; \\
$f$                    & friction coefficient; \\
$F_D$                    & drag force; \\
$F_G$                    & submerged weight; \\
$F_L$                    & lift force; \\
$g$                    & gravitational constant; \\
$h_b$                    & bed load layer thickness; \\
$k_s$                    & equivalent roughness height of Nikuradse; \\
$m$                    & total mass of the particle; \\
$q_b$                    & bed load sediment transport rate; \\
$R$                    & particle Reynolds number; \\
$R'^2$                    & correlation coefficient; \\
$s$                    & sediment relative density; \\
$T_*$                    & non-dimensional transport stage ($\equiv \tau_* / \tau_{*c}$); \\
$t$                    & time; \\
$t_{in}$                    & time required by the particle to reach $z_{in}$ (see splash function description); \\
$\Delta t$                    & numerical time step; \\
$U$                    & horizontal velocity component of the fluid; \\
$U_*$                    & shear flow velocity; \\
$U_{*c}$                    & critical shear velocity for sediment motion; \\
$u_s$                    & particle saltation streamwise velocity; \\
$\vec{V}$                    & ($\dot{x}$, $\dot{z}$) = particle velocity vector; \\
$\vec{V}_{in}$                    & $(\dot{x}, \dot{z})_{in}$ = particle velocity vector at collision; \\
$\vec{V}_{out}$                    & $(\dot{x}, \dot{z})_{out}$ = particle take-off velocity vector; \\
$V_r$                    & particle relative velocity evaluated at its centroid; \\
$V_{rB}$                    & relative velocity evaluated at the bottom of the particle; \\
$V_{rT}$                    & relative velocity evaluated at the top of the particle; \\
$x$                    & streamwise particle's centroid displacement; \\
$X$                    & any particle saltation characteristic (i.e. $\delta_s$, $\lambda_s$ or $u_s$); \\
$X_*$                    & any non-dimensional particle saltation characteristic; \\
$z$                    & bed-normal particle's centroid displacement; \\
$z_{in}$                    & collision line level; \\
$z_0$                    & zero-velocity bed level; \\
$\alpha_L$                    & lift coefficient for $F_{L_1}$; \\
$\alpha_m$                    & added mass coefficient; \\
$\beta$                    & bed slope; \\
$\delta_s$                    & particle saltation height; \\
$\theta_{out}$                    & take-off angle; \\
$\kappa$                    & von Kármán's constant; \\
$\lambda_s$                    & particle saltation length; \\
$\mu$                    & take-off angle mean value; \\
$\nu$                    & fluid kinematic viscosity; \\
$\rho$                    & fluid density; \\
$\rho_s$                    & sediment density; \\
$\sigma^2$                    & take-off angle variance; \\
$\tau_*$                    & non-dimensional bed shear stress; \\
$\tau_{*c}$                    & non-dimensional critical shear stress for sediment motion;   and \\
$\Phi $                    & non-dimensional bed load transport rate.

\end{longtable}

\end{document}